\documentclass[aps,prb,onecolumn,groupedaddress,showpacs]{revtex4} 

\usepackage{amssymb}
\usepackage{graphicx}
\usepackage{amsmath}
\usepackage{dsfont}
\usepackage{bbm}
\begin{document}

\title{Current and noise correlations in a double dot Cooper pair beam splitter}

\author{
	D.~Chevallier$^{1,2}$,
	J.~Rech$^{1}$,
	T.~Jonckheere$^{1}$, and
	T.~Martin$^{1,2}$
}

\affiliation{
	$^{1}$Centre de Physique Th\'eorique, 
	Case 907 Luminy, 13288 Marseille cedex 9, France
}

\affiliation{
	$^{2}$Universit\'e de la M\'edit\'erran\'ee, 
	13288 Marseille Cedex 9, France
}

\date{\today}

\begin{abstract}
We consider a double quantum dot coupled to two normal leads and one superconducting lead, modeling the Cooper pair beam splitter studied in two recent experiments. Starting from a microscopic Hamiltonian we derive a general expression for the branching current and the noise crossed correlations in terms of single and two-particle Green's function of the dot electrons. We then study numerically how these quantities depend on the energy configuration of the dots and the presence of direct tunneling between them, isolating the various processes which come into play. In absence of direct tunneling, the antisymmetric case (the two levels have opposite energies with respect to the superconducting chemical potential) optimizes the Crossed Andreev Reflection (CAR) process while the symmetric case (the two levels have the same energies) favors the Elastic Cotunneling (EC) process. Switching on the direct tunneling tends to suppress the CAR process, leading to negative noise crossed correlations over the whole voltage range for large enough direct tunneling.
\end{abstract}


\maketitle

\section{Introduction
\label{sec:introduction}}

At low temperatures, electron flow in mesoscopic systems bears analogies with the propagation of photons in quantum optics devices. The fermionic analog of the Hanbury Brown and Twiss experiment for photon is an example: negative current-current correlations demonstrate that the statistics of current carriers in microstuctures corresponds to a degenerate Fermi gas. Over the last two decades, the issue whether all fermionic systems should demonstrate antibunching has been addressed. It was predicted that if electrons are injected from a superconducting lead, into a  fork  consisting of two normal metal leads, then positive current-current crossed correlations could be observed \cite{anatram_datta,martin_ns,torres_martin,martin_torres}. This phenomenon, called Crossed Andreev Reflection (CAR), has since been interpreted as originating from the splitting of the constituent electrons of a Cooper pair from the superconductor into the two normal leads. Because this Cooper pair is a singlet pair of electrons, the two electrons which are injected into these leads should preserve their singlet nature, therefore providing a modern example of the non-local character of quantum mechanics in nanoelectronics.\cite{lesovik_martin_blatter,sukhorukov_recher_loss,recher_loss,deutscher_feinberg,belzig_brataas,zaikin_kalenkov}. Bell inequality measurements based on noise crossed correlations have been proposed to detect this entanglement.\cite{lesovik_martin_blatter,chtchelkatchev,chtchelkatchev_blatter_lesovik_martin,samuelson_buttiker,baowen_li}

\begin{figure}[ht]
	\centering
		\includegraphics[width=8cm]{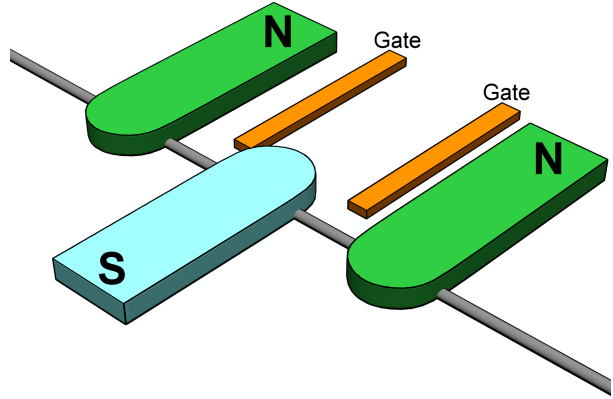}
	\caption{(color online) Double quantum dot coupled to normal/superconducting leads.}
\end{figure}

Over the last decade, CAR has been studied extensively for a variety of geometries. The early proposals pointed out that positive crossed correlations can be generated either by selecting the energies of the outgoing electrons, or alternatively by filtering their spins.\cite{lesovik_martin_blatter} Some theoretical works consider a superconductor connected in two separate locations to normal metal leads.\cite{bignon_houzet,feinberg,duhot_melin,melin_levy_yeyati,melin_freyn,falci_feinberg_hekking,melin_feinberg_europhys} Such works address the effect of the separation
between the normal metal leads on the CAR signal, and they specify which voltage configuration results in positive crossed correlations.  
Other theoretical works consider hybrid structures where the normal leads are taken to be ferromagnets or half metals\cite{deutscher_feinberg,melin_peysson,yamashita_takahashi_maekawa,melin_feinberg}, semiconductors\cite{brinkmann} and Luttinger liquids.\cite{bena,recher_loss_liquid} Recently the effect of Coulomb interaction has been studied in some of these systems.\cite{zaikin_golubev,yeyati_bergeret}   

On the experimental side, noise experiments with normal superconducting devices constitute a real challenge. A 
successfull experiment demonstrated that the Fano factor of a single NS junction is $2$, corresponding to the 
charge of a Cooper pair.\cite{mailly_sanquer} One of the intrinsic difficulties lies in achieving controlled, good quality contacts between the two arms and the superconductor.\cite{kontos_schonenberger_takayanagi} It has also been suggested that non-local effects in the current (Andreev drag) could be probed by placing two separate contacts to the superconductor. The first drag experiments were performed in this geometry \cite{beckman_prl,klapwijk}, but one true challenge remains that the signal for drag effects is very weak, and again that symmetric contacts are difficult to achieve.\cite{chandrasekar}

Two recent experiments\cite{hofstetter,herrmann} have provided evidence of Cooper pair splitting in devices consisting of a superconducting finger placed on the bulk of a single nanowire, whose both ends are connected to metallic leads (see Fig. 1). Because both the superconducting and the metallic leads are placed on top of the nanowire, two quantum dots are generated on both sides of the superconductor, 
and their energy levels can be controlled with the help of additional gates. This system constitutes a tunable Cooper pair beam splitter with a relatively good degree of symmetry. Differential conductance measurements showed appreciable non local signal which could be attributed to CAR. Some results also suggest that local interaction and direct tunneling between the dots may play an important role. In this work, we choose specifically to focus on the geometry of Refs. \onlinecite{hofstetter} and \onlinecite{herrmann}. The main goal of the present paper is to compute the branching currents of this device and their crossed correlations. These quantities depend strongly on the energy configuration of the dots as well as the presence of direct tunneling between them. In particular we study which parameters have to be optimized in order to achieve Cooper pair splitting. 
  
While exploring the above physics, we put some emphasis in presenting the formal aspects of the calculation, which are achieved starting from a microscopic Hamiltonian describing a BCS superconductor and normal leads, all coupled via tight binding tunnel Hamiltonian to the dots. Using a path integral formalism, the leads degrees of freedom are integrated out, which is equivalent to resumming the Dyson series in a perturbative Green's function approach. This allows to obtain a rather general formula for the current in terms of the Green's function of the dots. More importantly, we are also able to derive a general expression for the noise crossed correlations, in terms of a two particle dot Green's function. 

The outline of the paper is as follows. In Sec. \ref{sec:model}, we introduce the model for the Cooper pair splitter. The currents and the crossed correlations are computed in Sec. \ref{sec:current_and_noise}, and finally we apply our results to various situations (symmetric/antisymmetric case and with/without direct tunneling) in Sec. \ref{sec:results_and_discussion} to describe in which regime Elastic Cotunneling (EC) or CAR are optimized. We conclude in Sec. \ref{sec:conclusion}.   

\section{Model
\label{sec:model}}

In this section we introduce the Hamiltonian of the double dot system and derive the expression for the tunneling self-energy. To do so, and in order to clarify notations, we start by studying the single dot formalism.\cite{bayandin_martin,zazunov_egger_mora,cuevas} 

\subsection{Single dot formalism}

\subsubsection{Hamiltonian}

\begin{figure}[ht]
	\centering
		\includegraphics[width=6cm]{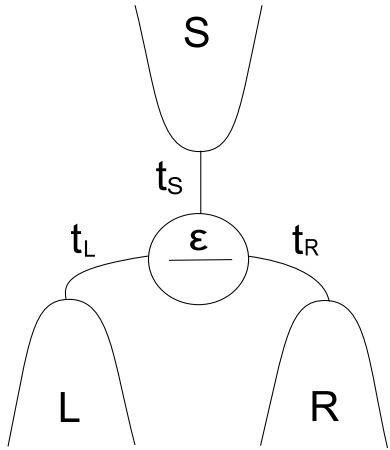}
	\caption{Single quantum dot coupled to normal/superconducting leads where $t_{L/R}$ ($t_S$) is respectively the tunneling amplitude between the dot and the normal (superconducting) lead.}
	\label{fig:systeme}
\end{figure}

We consider a quantum dot with a single level $\epsilon$, which is coupled via tunneling amplitudes $t_j$ ($j=L,R,S$) to two normal leads and one superconducting lead with superconducting gap $\Delta$ (see Fig. \ref{fig:systeme}). We label the applied bias voltage on the left (right) lead $V_L$
($V_R$), where $V_j$ is measured with respect to the chemical potential of the superconducting lead whose voltage is set at $V_S=0$. For simplicity we work with physical dimensions corresponding to $\hbar=1$ and $e=1$. The Hamiltonian of the total system reads
\begin{equation}
H = H_D + \sum_{j} H_j + H_T(t) ~,
\end{equation}
where the dot Hamiltonian is given by
\begin{equation}
H_{D} = \epsilon \, \sum_{\sigma = \uparrow, \downarrow}
d^\dagger_\sigma d_\sigma,
\end{equation}
and the lead Hamiltonians are expressed in terms of Nambu spinors
\begin{equation}
H_j = \sum_k \Psi^\dagger_{jk} \left(
\xi_k \, \sigma_z + \Delta_j \, \sigma_x \right) \Psi_{jk},
\end{equation}
with $\sigma_z$, $\sigma_x$ Pauli matrices in Nambu space and
\begin{equation}
\Psi_{jk} = \left(
\begin{array}{c}
\psi_{jk, \uparrow} \\
\psi^\dagger_{j(-k), \downarrow}
\end{array} \right)\:\textrm{and}\;
\xi_k = \frac{k^2}{2 m} - \mu ~.
\end{equation}\\
In the case of normal leads, $\Delta_j$ is zero. The tunneling Hamiltonian, which is responsible for transfer of electrons between the leads and the dot reads
\begin{equation}
H_T(t) = \sum_{jk} \,
\Psi^\dagger_{jk} \, {\cal T}_j(t) \, d + {\rm h.c.} ~,~~~
\label{H_T}
\end{equation}
where we introduce the Nambu spinor of the dot electrons 
\begin{equation}
d = \left(
\begin{array}{c}
d_\uparrow \\ d^\dagger_\downarrow
\end{array} \right).
\end{equation}
We include the voltage dependence in the tunneling term using the Peierls
substitution (performing a Gauge transformation in order to
represent the bias potential as a vector potential).
The tunneling amplitude thus reads
\begin{equation}
{\cal T}_j(t) = t_j \sigma_z \, e^{i \sigma_z \int V_j dt}.
\end{equation}
We now introduce the bare Green's functions of the dot (in the absence of tunneling)
\begin{equation}
\hat{G}_{0}^{ss'}(t,t')=-i\left\langle T_{C}\left\{d^s(t)d^{\dagger s'}(t')\right\}\right\rangle_0,
\end{equation}
where $T_C$ is the time ordering operator along the Keldysh contour, $s,s'$ labels the position of the times on this contour. The quantum mechanical
averaging is performed with respect to the Hamiltonian without
tunneling $\left\langle \dots \right\rangle_{0}=Z_{0}^{-1}\textrm{Tr} \left\{e^{-\beta
H_{0}}\dots\right\}$ with $Z_{0}=\textrm{Tr} \left\{ e^{-\beta H_{0}}\right\}$ and $H_0=H_D+\sum_j H_j$. The Green's function dressed by tunneling is written as
\begin{equation}
\hat{G}^{ss'}(t,t')=-i\left\langle T_{C}\left\{S(\infty)d^s(t)d^{\dagger s'}(t')\right\}\right\rangle_0,
\end{equation}  
where $S(\infty)$ is the evolution operator along the contour
\begin{equation}
S(\infty)=T_{C}\exp\left\{-i\int_{-\infty}^{+\infty}\!\!dt\,\sum\limits_{s=+,-}\tau_{z}^{ss}H_{T}^{s}(t)\right\},
\end{equation}
and $\tau_{z}$ is the $z$ Pauli matrix in the Keldysh space. 

\subsubsection{Averaging over the leads}

In this section we calculate the self-energy associated with the tunneling between the dot and the leads. Because the lead degrees of freedom appear quadratically in the total Hamiltonian, the evolution operator can straightforwardly be averaged over such leads 

\begin{equation}
\left\langle 
S(\infty)\right\rangle_{leads}=T_{C}\exp\left[-i\int_{C}\!\!dt_{1}dt_{2}\,
\hat{d}^{\dagger}(t_{1})\hat{\Sigma}_{T}(t_{1},t_{2})\hat{d}(t_{2})\right],
\end{equation}
where we introduce the spinor in the Nambu-Keldysh space
\begin{equation}
\hat{d}(t)=
\left(\begin{array}{c}
  d^{+}(t) \\
  d^{-}(t)
\end{array} \right).
\end{equation}
Here the self-energy associated with the tunneling takes the form
\begin{equation}
\hat{\Sigma}_{T}(t_1,t_2)=\sum\limits_{j=L,R,S}\hat{\Sigma}_{j}(t_{1},t_{2})=\sum\limits_{j=L,R,S}({\cal
T}_{j}^{\dagger}(t_1)\otimes\tau_{z})\hat{g}_{j}(t_{1}-t_{2})(\tau_{z}\otimes{\cal
T}_{j}(t_{2}))~,
\end{equation}
where $\hat{\Sigma}_T$, $\hat{g}_{j}$ are matrices in Nambu-Keldysh space,
and $\hat{g}_{j}(t-t^{\prime})=-i\sum_{k} \left\langle T_C \left\{\hat{\Psi}_{jk}(t) \hat{\Psi}_{jk}^{\dagger}(t')\right\}\right\rangle$ is the local Green's function of electrons on lead $j$. In the litterature\cite{cuevas}, they are typically given in the rotated Keldysh space,
\begin{equation}
\hat{g}^{RAK}\equiv\hat{L}\hat{\tau}_{z}\hat{g}\hat{L}^{-1}~,~~~\mbox{with}~~~
\hat{L}=\frac{1}{\sqrt{2}}\left(
\begin{array}{cc}
  1 & -1 \\
  1 & 1
\end{array}
\right)\otimes\mathds{1},
\end{equation}
where $\hat{\tau}_z=\tau_z\otimes\mathds{1}$, and $\mathds{1}$ is the unity matrix in the Nambu space. 
With this rotation the Green's function matrix can be simply expressed in terms
of the advanced, retarded and Keldysh components
\begin{equation}
\hat{g}^{RAK}=\left(
\begin{array}{cc}
  g^{R} & g^{K} \\
  0 & g^{A}
\end{array}
\right)~.
\end{equation}
The components of the leads Green's functions in the rotated base have the following form
\begin{equation}
\left\{
\begin{array}{lcl}
g_{j}^{R,A}(\omega)=\pi\nu(0)\frac{\omega\cdot{\mathds
1}+\Delta_{j}\cdot\sigma_{x}}{i\zeta_{\omega j}^{R,A}}~,  \\ \\ \\
g_{j}^{K}(\omega)=(1-2f_{\omega})\left(g_{j}^{R}(\omega)-g_{j}^{A}(\omega)\right)~,
\end{array}\right.
\end{equation}
\begin{equation}
\mbox{where}\ \  \zeta_{\omega j}^{R,A}=\left\{
\begin{array}{ll}
\pm~{\rm sign}(\omega)\sqrt{\omega^{2}-\Delta_{j}^{2}}~, &~~
|\omega|>\Delta_{j}, \\ \\
i\sqrt{\Delta_{j}^{2}-\omega^{2}}, &~~ |\omega|<\Delta_{j}~,
\end{array}
\right.
\end{equation}
with $\nu(0)$ the density of states of the normal lead at the Fermi level and $f_{\omega}$ the Fermi distribution. We rotate back in Keldysh space the matrix for the lead Green's function and we get the following formula for the self-energy $\hat{\Sigma}_j$ in time-domain
\begin{align}\label{self_energy_onedot}
\hat{\Sigma}_{j}(t_{1},t_{2})=\Gamma_{j}\int_{-\infty}^{\infty}\frac{d\omega}{2\pi}e^{-i\omega(t_{1}-t_{2})}
e^{-i\sigma_{z}V_{j}t_{1}}[\omega\cdot{\mathds{1}}-\Delta_{j}\cdot\sigma_{x}]e^{+i\sigma_{z}V_{j}t_{2}}\\\notag
\otimes\left[-\frac{\Theta(\Delta_{j}-|\omega|)}{\sqrt{\Delta_{j}^{2}-\omega^{2}}}\tau_{z}+i\,{\rm
sign}(\omega)\frac{\Theta(|\omega|-\Delta_{j})}{\sqrt{\omega^{2}-\Delta_{j}^{2}}}\left(\begin{array}{cc}
  2f_{\omega}-1 &~~ -2f_{\omega} \\
  +2f_{-\omega} &~~ 2f_{\omega}-1 \\
\end{array}\right)\right],
\end{align}

where $\Gamma_j=\pi\nu(0)\left|t_j\right|^2$ is the tunneling rate between the dot and the lead $j$.

\subsection{Double dot formalism}

\begin{figure}[ht]
		\includegraphics[width=6cm]{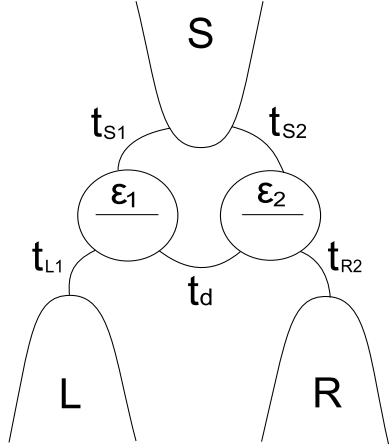}
		\caption{Double quantum dot coupled to normal/superconducting leads.}
		\label{fig:systeme2}
\end{figure}

For the remainder of this paper we focus on a system of two single-level quantum dots coupled to one superconducting lead and two normal leads (see Fig. \ref{fig:systeme2}). Such a system was studied experimentally in Refs. \onlinecite{hofstetter} and \onlinecite{herrmann}, using a carbon nanotube or a nanowire attached at both ends to normal metal leads with a superconducting electrode in the middle. 

The double dot formalism is an extension of the single dot one where there appears a new matrix structure in Dot space. The Hamiltonian of the total system reads

\begin{equation}
H=H_{D_1}+H_{D_2}+H_{D_1D_2}+\sum_{j}H_j+H_{T_1}(t)+H_{T_2}(t),
\end{equation}
where $H_{D_1}$ and $H_{D_2}$ are the Hamiltonians of the two dots
\begin{equation}
H_{D_\alpha}= \epsilon_\alpha \, \sum_{\sigma = \uparrow, \downarrow}
d^\dagger_{\alpha\sigma} d_{\alpha \sigma},
\end{equation}
and the tunneling between the dots is conveniently written using Nambu spinors and the tunneling amplitude $t_d$
\begin{equation}
H_{D_1D_2}=t_d d^{\dagger}_1\sigma_z d_2+h.c.~.
\end{equation}
These three terms can easily be combined under the following form
\begin{equation}
H_D=\tilde{d}^{\dagger}\begin{pmatrix}
\epsilon_1	&t_d\\
t_d						&\epsilon_2
\end{pmatrix}\otimes\sigma_z\tilde{d},
\end{equation}
where $\tilde{d}^{\dagger}=\left( 
d^{\dagger}_1\;d^{\dagger}_2
\right)$ is a spinor in the Nambu-Dot space.
The tunneling between the dot $\alpha$ and the leads reads
\begin{equation}
H_{T_\alpha}(t) = \sum_{jk} \,
\Psi^\dagger_{jk} \, {\cal T}_{j\alpha}(t) \, d_\alpha + h.c.~, 
\end{equation}
where ${\cal T}_{j\alpha}(t)=t_{j\alpha} \sigma_z \, e^{i \sigma_z \int V_j dt}$, and $t_{j\alpha}$ corresponds to the tunneling amplitude between the dot $\alpha$ and the lead $j$. This Hamiltonian corresponds to a system where every dot is coupled to every lead. In order to reproduce the system of Fig. \ref{fig:systeme2}, it is necessary to set to zero the tunneling amplitudes $t_{L2}$ and $t_{R1}$ coupling the dot $2$ with the left lead and the dot $1$ with the right lead.

Now the Green's function of the dot electrons has a new form
\begin{equation}\label{green_ddot}
\check{G}_{\alpha\alpha'}^{ss'}(t,t')=-i\left\langle T_C\left\{d^{s}_{\alpha}(t)d^{\dagger s'}_{\alpha'}(t')\right\}\right\rangle,
\end{equation}
with $\alpha,\alpha'$ corresponding to the dot index. As in the previous section the self-energy associated with the tunneling between the dots and the leads is calculated by averaging the evolution operator over the leads degrees of freedom
\begin{equation}
\left\langle S(\infty)\right\rangle_{leads}=T_{C} \exp\left[-i \int^{+\infty}_{-\infty}dt_1dt_2 \check{d}^{\dagger}(t_1)\check{\Sigma}_T(t_1,t_2)\check{d}(t_2)\right],
\end{equation}
where $\check{d}(t)=
\left(\begin{array}{c}
  \tilde{d}^{+}(t) \\
  \tilde{d}^{-}(t)
\end{array} \right)$ is a spinor in the Nambu-Dot-Keldysh space and the self-energy is given by
\begin{equation}\label{self_energy_twodot}
\check{\Sigma}_T(t_1-t_2)=\sum_{j}\begin{pmatrix}
({\cal
T}_{j1}^{\dagger}(t_1)\otimes\tau_{z})\hat{g}_{j}(t_{1}-t_{2})(\tau_{z}\otimes{\cal
T}_{j1}(t_{2}))	&({\cal
T}_{j1}^{\dagger}(t_1)\otimes\tau_{z})\hat{g}_{j}(t_{1}-t_{2})(\tau_{z}\otimes{\cal
T}_{j2}(t_{2}))\\
({\cal
T}_{j2}^{\dagger}(t_1)\otimes\tau_{z})\hat{g}_{j}(t_{1}-t_{2})(\tau_{z}\otimes{\cal
T}_{j1}(t_{2}))	&({\cal
T}_{j2}^{\dagger}(t_1)\otimes\tau_{z})\hat{g}_{j}(t_{1}-t_{2})(\tau_{z}\otimes{\cal
T}_{j2}(t_{2}))
\end{pmatrix},
\end{equation}
where $\hat{g}_{j}$ and $\check{\Sigma}_T$ are matrices in Nambu-Keldysh and Nambu-Dot-Keldysh space respectively. Each element $\hat{\Sigma}_{j\alpha\beta}(t_1-t_2)$ of the self-energy matrix (\ref{self_energy_twodot}) in the Dot space can be obtained from  Eq.~(\ref{self_energy_onedot}) by replacing $\Gamma_j$ with $\Gamma_{j\alpha\beta}=\pi\nu(0)t^{*}_{j\alpha}t_{j\beta}$.

\section{Currents and Crossed correlations}
\label{sec:current_and_noise}

In this section we derive the currents between the two dots and the various leads as well as their crossed correlations using the dot electrons Green's function and the tunneling self-energy calculated previously.

\subsection{Currents}

The current from the dot $\alpha$ into the lead $j$ reads 
\begin{equation}
I_{j\alpha}(t)=i\sum_k \Psi^{\dagger}_{j k}\sigma_z{\cal
T}_{j\alpha}(t)d_{\alpha} + h.c.~.
\end{equation}
The average current does not depend on which branch of the Keldysh contour the time is chosen, and therefore can be expressed as $\left\langle I_{j\alpha}\right\rangle=\left\langle I_{j\alpha}(t^{+})+I_{j\alpha}(t^-)\right\rangle/2$ where $t^{\pm}$ is the time on the upper/lower branch of the contour. In order to compute these it is convenient to introduce counting fields $\eta_{j\alpha}(t)$ which appear in the tunneling amplitudes as ${\cal
T}_{j\alpha}(t)\rightarrow {\cal
T}_{j\alpha}(t)e^{i\tau_z\otimes \sigma_z \eta_{j\alpha}(t)/2}$. The average current from the dot $\alpha$ into the lead $j$ can then be calculated as the first derivative of the Keldysh partition function
\begin{equation}
\left\langle I_{j\alpha}\right\rangle=i\frac{1}{Z[0]}\left.\frac{\delta Z\left[\eta\right]}{\delta \eta_{j\alpha}(t)}\right|_{\eta=0},
\end{equation}
where $Z[\eta]=\left\langle S(\infty,\eta)\right\rangle_0$ and $S(\infty,\eta)$ is the evolution operator in which the counting fields were introduced. After performing the derivative we obtain the following result for the current
\begin{align}
\left\langle I_{j\alpha}\right\rangle=&\frac{1}{2}\textrm{Tr}\left\{(\tau_z\otimes \sigma_z)\int^{+\infty}_{-\infty}dt'\left(\check{G}(t,t')\check{\Sigma}_{j}(t',t)-\check{\Sigma}_{j}(t,t')\check{G}(t',t)\right)_{\alpha\alpha}\right\},
\end{align}
where "$\textrm{Tr}$" corresponds to the trace in Nambu-Keldysh space. Going to the rotated Keldysh space, the average current can be re-expressed in terms of the advanced, retarded and Keldysh components as
\begin{align}
\left\langle I_{j\alpha}\right\rangle=&\frac{1}{2}\textrm{tr}\left\{\sigma_z \int^{+\infty}_{-\infty}dt'\left(\tilde{G}^R(t,t')\tilde{\Sigma}^K_{j}(t',t)+\tilde{G}^K(t,t')\tilde{\Sigma}^{A}_{j}(t',t)-\tilde{\Sigma}^R_{j}(t,t')\tilde{G}^K(t',t)-\tilde{\Sigma}^K_{j}(t,t')\tilde{G}^{A}(t',t)\right)_{\alpha\alpha}\right\},
\end{align}
where "$\textrm{tr}$" corresponds to the trace in Nambu space. While the self-energy $\tilde{\Sigma}_j$ can be obtained from the results of the previous section, the Green's function $\tilde{G}$ remains to be determined. To do this we write the Dyson equation in the frequency domain and we obtain the various components of $\tilde{G}$ in the rotated Keldysh space
\begin{align}\label{dyson_equation}
\tilde{G}^{R/A}(\omega)^{-1}&=\tilde{G}^{R/A}_{0}(\omega)^{-1}-\tilde{\Sigma}^{R/A}_{T}(\omega),\\
\tilde{G}^{K}(\omega)&=\tilde{G}^{K}_{0}(\omega)+\tilde{G}^{R}(\omega)\tilde{\Sigma}^{K}_{T}(\omega)\tilde{G}^{A}(\omega),\label{dyson_equation2}
\end{align}
with
\begin{align}
\tilde{G}^{R/A}_{0}(\omega)^{-1}&=\begin{pmatrix}
\omega \mathds{1}-\epsilon_{1}\sigma_z	&-t_{d}\sigma_z\\
-t_{d}\sigma_z	&\omega \mathds{1}-\epsilon_{2}\sigma_z
\end{pmatrix},\\
\tilde{G}^{K}_{0}(\omega)&=0.
\end{align}
In terms of these Fourier transformed functions, the current can be rewritten as 
\begin{eqnarray}\label{current}
\left\langle I_{j\alpha}\right\rangle={\rm{}tr}\left\{\sigma_{z}\int\limits_{-\infty}^{+\infty}\frac{d\omega}{2\pi}
{\rm{}Re}\left[\left(\tilde{G}^{R}(\omega)\tilde{\Sigma}_{j}^{K}(\omega)+\tilde{G}^{K}(\omega)\tilde{\Sigma}_{j}^{A}(\omega)\right)_{\alpha\alpha}\right]\right\}~,
\end{eqnarray}
where we used that the Keldysh components are anti-Hermitian while the advanced and retarded components are the Hermitian conjugate 
of one another. We can now calculate the current using the Dyson equations (\ref{dyson_equation})-(\ref{dyson_equation2}) and the components of the tunneling self-energy obtained in the previous section which read in the frequency domain
\begin{eqnarray}
\tilde{\Sigma}_{R}^{A/R}(\omega)&=&\pm{}i\begin{pmatrix}
0	&0\\
0	&\Gamma_{R22}
\end{pmatrix}\otimes
\mathds{1}~,\\
\tilde{\Sigma}_{L}^{A/R}(\omega)&=&\pm{}i\begin{pmatrix}
\Gamma_{L11}	&0\\
0	&0
\end{pmatrix}\otimes
\mathds{1},\\
\tilde{\Sigma}_{S}^{A/R}(\omega)&=&X_{S}^{A/R}(\omega)\begin{pmatrix}
\Gamma_{S11}	&\Gamma_{S12}\\
\Gamma_{S21}	&\Gamma_{S22}
\end{pmatrix}\otimes\left(\begin{array}{cc}
  1 & -\frac{\Delta}{\omega} \\
  -\frac{\Delta}{\omega} & 1
\end{array}\right)~,\\
\tilde{\Sigma}_{R}^{K}(\omega)&=&-2i\begin{pmatrix}
0	&0\\
0	&\Gamma_{R22}
\end{pmatrix}\otimes\left(\begin{array}{cc}
  {\rm{}tanh}\left(\frac{\beta(\omega-V_{R})}{2}\right) & 0 \\
  0 & {\rm{}tanh}\left(\frac{\beta(\omega+V_{R})}{2}\right)
\end{array}\right)~,\\
\tilde{\Sigma}_{L}^{K}(\omega)&=&-2i\begin{pmatrix}
\Gamma_{L11}	&0\\
0	&0
\end{pmatrix}\otimes\left(\begin{array}{cc}
  {\rm{}tanh}\left(\frac{\beta(\omega-V_{L})}{2}\right) & 0 \\
  0 & {\rm{}tanh}\left(\frac{\beta(\omega+V_{L})}{2}\right)
\end{array}\right)~,\\
\tilde{\Sigma}_{S}^{K}(\omega)&=&X_{S}^{K}(\omega)\begin{pmatrix}
\Gamma_{S11}	&\Gamma_{S12}\\
\Gamma_{S21}	&\Gamma_{S22}
\end{pmatrix}\otimes\left(\begin{array}{cc}
  1 & -\frac{\Delta}{\omega} \\
  -\frac{\Delta}{\omega} & 1
\end{array}\right)~,
\end{eqnarray}
where we focused on the case $V_{S}=0$ allowing us to simplify the expression for the self-energy $\tilde{\Sigma}_S$, and introduced
\begin{equation}
X^{A/R}_{S}(\omega)=-\frac{\Theta(\Delta-\left|\omega\right|)\omega}{\sqrt{\Delta^{2}-\omega^{2}}}\pm i\frac{\Theta(\left|\omega\right|-\Delta)\left|\omega\right|}{\sqrt{\omega^{2}-\Delta^{2}}},
\end{equation}
\begin{equation}
X^{K}_{S}(\omega)=-2i\frac{\Theta(\left|\omega\right|-\Delta)\left|\omega\right|}{\sqrt{\omega^{2}-\Delta^{2}}}\textrm{tanh}\left(\frac{\beta\omega}{2}\right).
\end{equation}

\subsection{Crossed correlations}

We follow a similar approach to the one developed to calculate the currents. Here however the crossed correlations involve two current operators evaluated at different times which in general do not commute. Usually the measurement procedure dictates which combination of the current-current correlators is involved in the expression of the measured noise.\cite{lesovik_loosen,gavish,aguado_kouvenhoven,deblock} For this reason, we need to compute the unsymmetrized correlator, which can subsequently be manipulated to be applied to a desired measurement procedure. 

We introduce a partition function $Z[\eta]=\left\langle S(\infty,\eta)\right\rangle_0$ depending on a counting field $\eta_{j\alpha s}(t)$ where $j=L,R,S$ corresponds to the different leads, $\alpha=1,2$ to the two dots and $s=\pm$ to the branches of the Keldysh contour. This new counting field enters in the tunneling amplitudes as
\begin{equation}
{\cal
T}_{j\alpha}\rightarrow{\cal
T}_{j\alpha}e^{\sum_{s}i\pi_{s}\otimes\sigma_z\eta_{j\alpha s}(t)},
\end{equation}
where we introduced $\pi$-matrices in Keldysh space
\begin{equation}
\pi_{+}=\begin{pmatrix}
1	&0\\
0	&0
\end{pmatrix},\pi_{-}=
\begin{pmatrix}
0	&0\\
0	&-1
\end{pmatrix}.
\end{equation}
Now the crossed correlation can be calculated as a second derivative of this partition function over the counting fields
\begin{equation}
\left\langle I^{-}_{i\alpha}(t)I^{+}_{j\beta}(t')\right\rangle=-\frac{1}{Z[0]}\left.\frac{\delta^2 Z[\eta]}{\delta\eta_{i\alpha-}(t)\delta\eta_{j\beta+}(t')}\right|_{\eta\rightarrow0}~.
\end{equation}
Performing the average over the leads allows us to write the current-current correlation function
\begin{align}\label{correlationtwopart}
&\left\langle I^{-}_{i\alpha}(t)I^{+}_{j\beta}(t')\right\rangle=\int dt_1 dt_2 \sum_{\gamma\delta ss'} \sum_{\sigma_1\sigma_2\sigma_1'\sigma_2'}\sigma_1\sigma_1'\notag\\
&\times\left(\check{\Sigma}^{-s}_{i,\alpha\gamma\sigma_1\sigma_2}(t,t_1)\check{\Sigma}^{+s'}_{j,\beta\delta\sigma_1'\sigma_2'}(t',t_2)\check{K}^{ss'-+}_{\begin{subarray}{1}
\gamma\delta\alpha\beta\\ \sigma_2\sigma_2'\sigma_1\sigma_1'
\end{subarray}}(t_1,t_2,t,t')\right.-\check{\Sigma}^{-s}_{i,\alpha\gamma\sigma_1\sigma_2}(t,t_1)\check{\Sigma}^{s'+}_{j,\delta\beta\sigma_2'\sigma_1'}(t_2,t')\check{K}^{s+-s'}_{\begin{subarray}{1}
\gamma\beta\alpha\delta\\ \sigma_2\sigma_1'\sigma_1\sigma_2'
\end{subarray}}(t_1,t',t,t_2)\notag\\
&-\check{\Sigma}^{s-}_{i,\gamma\alpha\sigma_2\sigma_1}(t_1,t)\check{\Sigma}^{+s'}_{j,\beta\delta\sigma'\sigma_2'}(t',t_2)\check{K}^{-s's+}_{\begin{subarray}{1}
\alpha\delta\gamma\beta\\ \sigma_1\sigma_2'\sigma_2\sigma_1'
\end{subarray}}(t,t_2,t_1,t')\left.+\check{\Sigma}^{s-}_{i,\gamma\alpha\sigma_2\sigma_1}(t_1,t)\check{\Sigma}^{s'+}_{j,\delta\beta\sigma_2'\sigma_1'}(t_2,t')\check{K}^{-+ss'}_{\begin{subarray}{1}
\alpha\beta\gamma\delta\\ \sigma_1\sigma_1'\sigma_2\sigma_2'
\end{subarray}}(t,t',t_1,t_2)\right)~,
\end{align}
where 
\begin{equation}
\check{K}^{s_1s_2s_3s_4}_{\begin{subarray}{1}
\alpha_1\alpha_2\alpha_3\alpha_4\\ \sigma_1\sigma_2\sigma_3\sigma_4
\end{subarray}}(t_1,t_2,t_3,t_4)=-\left\langle T_C\left\{{\check{d}^{s_1}_{\alpha_1\sigma_1}(t}_1)\check{d}^{s_2}_{\alpha_2\sigma_2}(t_2)\check{d}_{\alpha_3\sigma_3}^{\dagger s_3}(t_3)\check{d}_{\alpha_4\sigma_4}^{\dagger s_4}(t_4)\right\}\right\rangle
\end{equation}
is the two particle Green's function of the dots electrons and $\check{\Sigma}^{s_1s_2}_{j,\alpha_1\alpha_2\sigma_1\sigma_2}(t_1,t_2)$ is the matrix element of the tunneling self-energy associated with the lead $j$. In the general case where Coulomb interaction is present on the dots, the two-particle Green's function can be expressed in terms of the dressed single particle Green's function and the full vertex function. However in our non-interacting case the two-particle Green's function reduces to 
\begin{align}\label{twopart}
\check{K}^{s_1s_2s_3s_4}_{\begin{subarray}{1}
\alpha_1\alpha_2\alpha_3\alpha_4\\ \sigma_1\sigma_2\sigma_3\sigma_4
\end{subarray}}(t_1,t_2,t_3,t_4)=\check{G}^{s_1s_4}_{\alpha_1\alpha_4\sigma_1\sigma_4}(t_1,t_4)\check{G}^{s_2s_3}_{\alpha_2\alpha_3\sigma_2\sigma_3}(t_2,t_3)-\check{G}^{s_1s_3}_{\alpha_1\alpha_3\sigma_1\sigma_3}(t_1,t_3)\check{G}^{s_2s_4}_{\alpha_2\alpha_4\sigma_2\sigma_4}(t_2,t_4)~,
\end{align}
where $G^{s_1s_2}_{\alpha_1\alpha_2\sigma_1\sigma_2}(t_1,t_2)$ is the $\sigma_1\sigma_2$ matrix element of the single particle Green's function introduced in Eq. (\ref{green_ddot}). Substituting Eq. (\ref{twopart}) into Eq. (\ref{correlationtwopart}) we obtain the irreducible part of the current-current correlator
\begin{align}
S_{i\alpha,j\beta}(t,t')&=\left\langle I_{i\alpha}(t)I_{j\beta}(t')\right\rangle-\left\langle I_{i\alpha}(t)\right\rangle\left\langle I_{j\beta}(t')\right\rangle\notag\\
=&-\int^{+\infty}_{-\infty}dt_1dt_2 \textrm{Tr}\left\{(\pi_{-}\otimes\sigma_z)(\check{\Sigma}_i(t,t_1)\check{G}(t_1,t'))_{\alpha\beta}(\pi_{+}\otimes\sigma_z)(\check{\Sigma}_j(t',t_2)\check{G}(t_2,t))_{\beta\alpha}\right.\notag\\
&+(\pi_{-}\otimes\sigma_z)(\check{G}(t,t_1)\check{\Sigma}_j(t_1,t'))_{\alpha\beta}(\pi_{+}\otimes\sigma_z)(\check{G}(t',t_2)\check{\Sigma}_i(t_2,t))_{\beta\alpha}\notag\\
&-(\pi_{-}\otimes\sigma_z)(\check{\Sigma}_i(t,t_1)\check{G}(t_1,t_2)\check{\Sigma}_j(t_2,t'))_{\alpha\beta}(\pi_{+}\otimes\sigma_z)\check{G}_{\beta\alpha}(t',t)\notag\\
&-\left.(\pi_{-}\otimes\sigma_z)\check{G}_{\alpha\beta}(t,t')(\pi_{+}\otimes\sigma_z)(\check{\Sigma}_j(t',t_1)\check{G}(t_1,t_2)\check{\Sigma}_i(t_2,t))_{\beta\alpha}\right\}~.
\end{align}
Performing a rotation in Keldysh space, the irreducible part of the current-current correlation function can be expressed in terms of the advanced, retarded and Keldysh components as
\begin{align}
&S_{i\alpha,j\beta}(t,t')=-\frac{1}{2}{\rm{}Re}\int\limits_{-\infty}^{+\infty}\int\limits_{-\infty}^{+\infty}dt_{1}dt_{2}\notag\\
&\times\textrm{tr}\left\{\sigma_{z}\left(\tilde{\Sigma}_{i}^{K}\tilde{G}^{A}+\tilde{\Sigma}_{i}^{R}\tilde{G}^{K}-\tilde{\Sigma}_{i}^{A}\tilde{G}^{A}+\tilde{\Sigma}_{i}^{R}\tilde{G}^{R}\right)^{\alpha\beta}_{(t,t_{1})\circ(t_{1},t^{\prime})}
\sigma_{z}\left(\tilde{\Sigma}_{j}^{K}\tilde{G}^{A}+\tilde{\Sigma}_{j}^{R}\tilde{G}^{K}+\tilde{\Sigma}_{j}^{A}\tilde{G}^{A}-\tilde{\Sigma}_{j}^{R}\tilde{G}^{R}\right)^{\beta\alpha}_{(t^{\prime},t_{2})\circ(t_{2},t)}\right.\notag\\
&\left.-\sigma_{z}\left(\tilde{\Sigma}_{i}^{R}\tilde{G}^{R}\tilde{\Sigma}_{j}^{K}+\tilde{\Sigma}_{i}^{K}\tilde{G}^{A}\tilde{\Sigma}_{j}^{A}+\tilde{\Sigma}_{i}^{R}\tilde{G}^{K}\tilde{\Sigma}_{j}^{A}
-\tilde{\Sigma}_{i}^{A}\tilde{G}^{A}\tilde{\Sigma}_{j}^{A}+\tilde{\Sigma}_{i}^{R}\tilde{G}^{R}\tilde{\Sigma}_{j}^{R}\right)^{\alpha\beta}_{(t,t_{1})\circ(t_{1},t_{2})\circ(t_{2},t^{\prime})}
\sigma_{z}\left(\tilde{G}^{K}+\tilde{G}^{A}-\tilde{G}^{R}\right)^{\beta\alpha}_{(t^{\prime},t)}\right\}~.
\nonumber
\end{align}
Finally we take the Fourier transform of this expression and we obtain the irreducible part of the frequency dependent current-current correlation
\begin{align}\label{noise}
S_{i\alpha,j\beta}(\omega)&=-\frac{1}{2}{\rm{}Re}\!\int\limits_{-\infty}^{+\infty}\frac{d\omega^{\prime}}{2\pi}{\rm
tr}\left\{
\sigma_{z}\left(\tilde{\Sigma}_{i}^{K}\tilde{G}^{A}+\tilde{\Sigma}_{i}^{R}\tilde{G}^{K}-\tilde{\Sigma}_{i}^{A}\tilde{G}^{A}+\tilde{\Sigma}_{i}^{R}\tilde{G}^{R}\right)^{\alpha\beta}_{\omega^{\prime}}
\sigma_{z}\left(\tilde{\Sigma}_{j}^{K}\tilde{G}^{A}+\tilde{\Sigma}_{j}^{R}\tilde{G}^{K}+\tilde{\Sigma}_{j}^{A}\tilde{G}^{A}-\tilde{\Sigma}_{j}^{R}\tilde{G}^{R}\right)^{\beta\alpha}_{\omega+\omega^{\prime}}\right.\notag\\
&\left.-\sigma_{z}\left(\tilde{\Sigma}_{i}^{R}\tilde{G}^{R}\tilde{\Sigma}_{j}^{K}+\tilde{\Sigma}_{i}^{K}\tilde{G}^{A}\tilde{\Sigma}_{j}^{A}+\tilde{\Sigma}_{i}^{R}\tilde{G}^{K}\tilde{\Sigma}_{j}^{A}
-\tilde{\Sigma}_{i}^{A}\tilde{G}^{A}\tilde{\Sigma}_{j}^{A}+\tilde{\Sigma}_{i}^{R}\tilde{G}^{R}\tilde{\Sigma}_{j}^{R}\right)^{\alpha\beta}_{\omega^{\prime}}
\sigma_{z}\left(\tilde{G}^{K}+\tilde{G}^{A}-\tilde{G}^{R}\right)^{\beta\alpha}_{\omega+\omega^{\prime}}\right\}.
\end{align}

In the following section we compute numerically the current (\ref{current}) and the crossed correlation (\ref{noise}) for various regimes and we comment on the results.

\section{Results and discussion}
\label{sec:results_and_discussion}

\subsection{Transport without tunneling between the dots}
\label{without_tunneling}

To begin we present the three dominant processes which can occur in such a system. Then we study two situations: the anti-symmetric case  
when the energy of the two dots have the opposite position 
(with respect to the chemical potential of the superconducting lead at $V_S=0$)
and the symmetric case when the energy levels of the two dots are the same. 
All energy scales in this section are in units of $\Delta$ and in the subgap regime. In all the following 
results we focus on the low temperature regime ($\beta \gg 1/\Delta$) as the only effect of temperature is 
to smooth the signal.

\subsubsection{Dominant electron transfer processes}

\begin{figure}[ht]
	\centering
		\includegraphics[width=5cm]{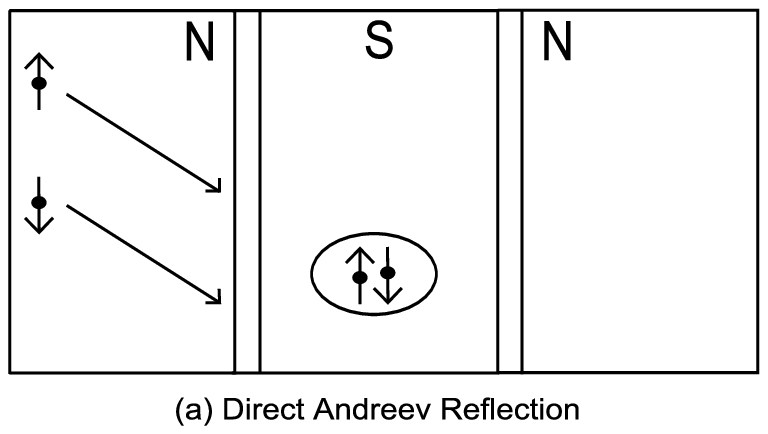}
		\includegraphics[width=5cm]{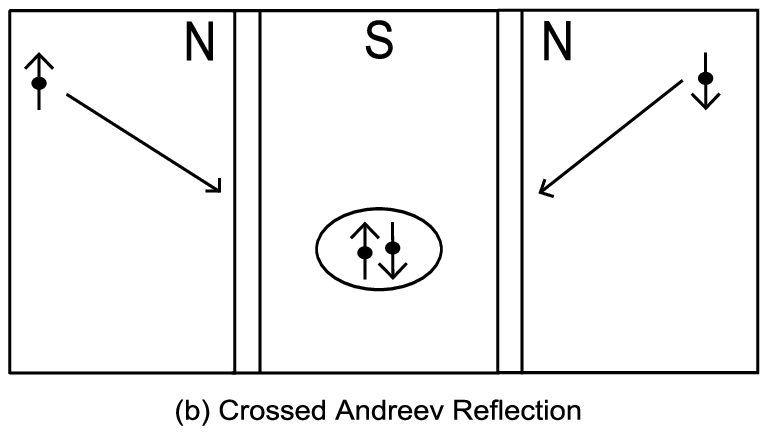}
		\includegraphics[width=5cm]{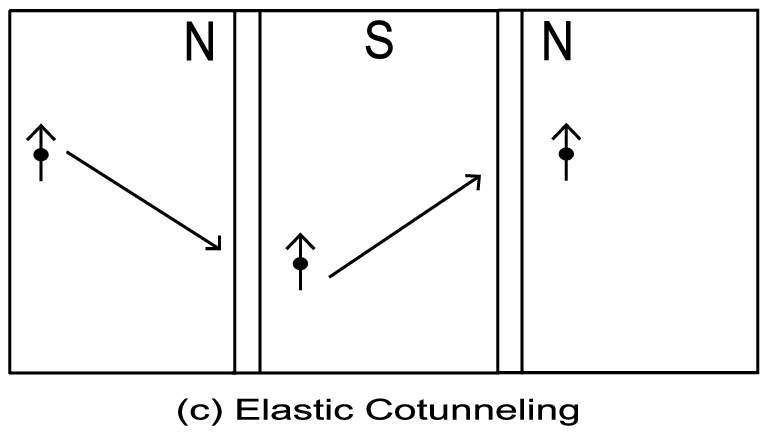}
		\caption{Dominant electron transfer processes}
	\label{fig:DAR}
\end{figure}

\textbf{Direct Andreev Reflection (DAR)} (see Fig. \ref{fig:DAR}a)
This process involves an electron which is incident from a normal lead at energies less than the superconducting energy gap. The incident electron forms a Cooper pair in the superconductor with the retroreflection of a hole of opposite spin and momentum to this incident electron. This is equivalent to two electrons with opposite spin and momentum from the same lead forming a Cooper pair in the superconductor. The inverse process corresponds to the destruction of a Cooper pair in the superconductor with two electrons propagating in the same lead. \\

\textbf{Crossed Andreev Reflection (CAR)} (see Fig. \ref{fig:DAR}b)
Crossed Andreev reflection occurs when two spatially separated normal leads form two separate junctions with a superconductor
(the separation between the two normal leads is assumed to be smaller than the BCS superconducting coherence length). In such a device, retroreflection of the hole from an Andreev reflection process, resulting from an incident electron at energies less than the superconducting gap at one lead, occurs in the second spatially separated normal lead with the same charge transfer as in a normal Andreev Reflection (AR) process to a Cooper pair in the superconductor. For CAR to occur, electrons of opposite spin and energies must exist at each normal lead (so as to form the pair in the superconductor). The inverse process corresponds to the destruction of a Cooper pair with two electrons propagating in opposite leads. \\

\textbf{Elastic Cotunneling (EC)} (see Fig. \ref{fig:DAR}c)
Elastic cotunelling is the quantum mechanical tunneling of electrons between the normal leads via an intermediate state in the superconductor. It does not lead to Cooper pair creation or annihilation in the superconductor.\\

According to the configuration of our system one or several of these processes occur. The goal is to study which configuration facilitates which process.

\subsubsection{Anti-symmetric case}
\label{anti_symmetric_no_tunneling}

\begin{figure}[ht]
	\centering
		\includegraphics[width=8cm]{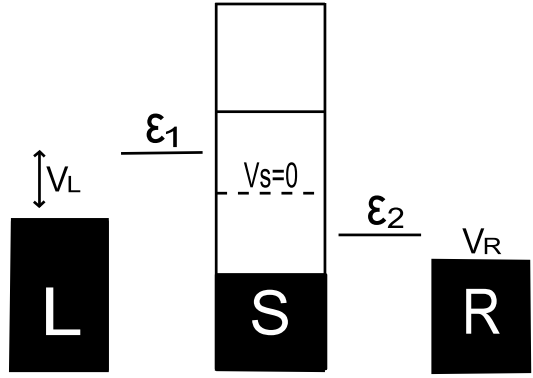}
	\caption{Anti-symmmetric case: the energy levels of the dots are opposite with respect to the chemical potential of the superconducting lead whose voltage is set at $V_S=0$. The voltage of the right lead $V_R$ is fixed and we vary the voltage of the left one $V_L$.}
	\label{fig:AntiSym}
\end{figure}

In this section we focus on the geometry where the two dots levels are opposite with respect to the superconducting chemical potential. We fix the voltage of the right lead and we vary the voltage of the left one (Fig. \ref{fig:AntiSym}).

\begin{figure}[ht]
	\centering
		\includegraphics[width=10cm]{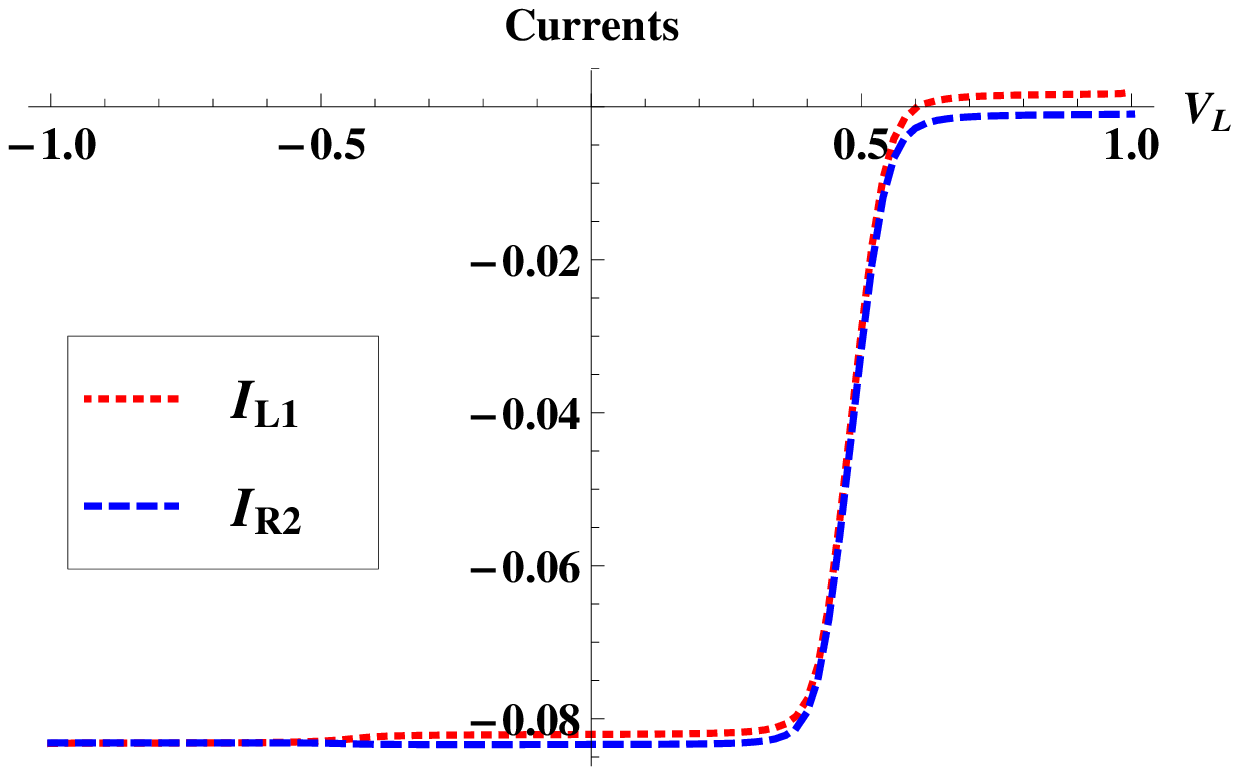}
	\caption{(Color online) Currents (arbitrary units) as a function of the voltage of the left normal lead $V_L$ for $\epsilon_1=0.5$, $\epsilon_2=-0.5$, $V_R=-0.7$, $\beta=100$, $t_{L1}=t_{R2}=t_{S1}=t_{S2}=0.2$ and $t_{L2}=t_{R1}=0$.}
	\label{fig:Courant2}
\end{figure}

\begin{figure}[ht]
	\centering
		\includegraphics[width=10cm]{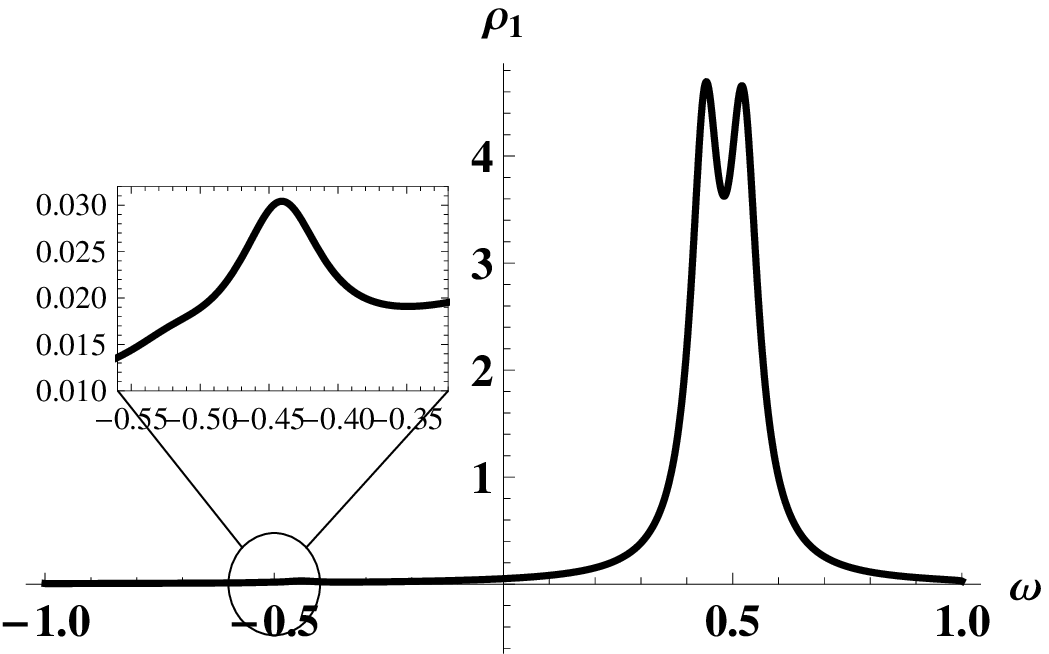}
	\caption{Density of states of the dot 1 (arbitrary units) for the same setup as in Fig. \ref{fig:Courant2}.}
	\label{fig:Density0anti}
\end{figure}

In Fig. \ref{fig:Courant2} we plot the current which flows in the two normal leads ($I_{L1}$ is the current between the dot $1$ and the left normal lead and $I_{R2}$ is the current between the dot $2$ and the right normal lead). We see that for voltages below 
$\epsilon_1$ (the energy level of the first dot) both currents have the same sign (with our notation currents which enter the leads are positive) and essentially the same amplitude. 
(strictly speaking, we observe a small deviation between the two currents around $V_L=\epsilon_2$).
Above $\epsilon_1$ the currents have a reduced but comparable amplitude but their signs are opposite.

When the voltage of the left lead is smaller than $\epsilon_2$,
the dominant process is the Crossed Andreev Reflection (CAR) in the sense that a Cooper pair from the superconducting lead is split and its constituent electrons are injected in the two leads. Strictly speaking, below $\epsilon_2$ Direct Andreev Reflection (DAR)
gives also a contribution, albeit a minor one. This can be understood by looking at the density of states of the dots:
\begin{equation}
\rho_{\alpha}=\frac{1}{\pi}\textrm{Im}(\tilde{G}^{A})_{\uparrow\uparrow\alpha\alpha}~,
\end{equation}
where $\alpha=1,2$. $\rho_{1}$ is plotted in Fig. \ref{fig:Density0anti}. It contains a sharp double peak at $\epsilon_1$, and a much 
smaller peak at $-\epsilon_1=\epsilon_2$ which originates from the proximity effect of the superconducting lead. In this situation $\rho_{2}$ (not shown) is exactly the symmetric of $\rho_{1}$ with respect to the $\omega=0$ axis.
As electrons can tunnel through these two resonances in the same lead, this explains the presence of a weak DAR process.  

The CAR and DAR processes thus give currents ($I_{L1}$ and $I_{R2}$) with the same sign because electrons injected from the superconductor only end up in the two normal leads. Elastic Cotunneling (EC) does not give any contribution because the voltages of each normal leads are not large enough 
to allow quasiparticle injection in the superconductor. 
Note that the amplitude of the generated currents by CAR is dominant because this configuration is optimal for the process of resonant electron transfer through the (large) peaks of the density of states $\rho_{1}$ and $\rho_{2}$ (at opposite energies).

As soon as $V_L>-\epsilon_1$, the two currents deviate slightly from each other because EC comes into play. Indeed, an 
electron from the left lead can tunnel into the superconductor via the (small) resonance in $\rho_{1}$. 
It may continue its way to dot 2 using the (large) resonance in $\rho_{2}$ which is located at the same energy. 
Note at the same time that DAR processes continue to operate, but only for the right lead. 
This explains why $I_{L1}>I_{R2}$ in the region $[ -\epsilon_1, \epsilon_1 ]$. 

When the voltage of the left lead is larger than $\epsilon_1$, the CAR process is strongly suppressed because 
$V_L$ is placed above all resonant levels of $\rho_{1}$ and $\rho_{2}$. Electrons injected from the superconductor
are prohibited from entering the normal left lead. The only allowed processes are thus EC and DAR. DAR process from the left lead injects Cooper pairs in the superconductor, while they destroy Cooper pairs which are injected as electrons 
in the right lead. In addition, EC contributes to transfer electrons from the left lead to the right lead via the superconductor.
The amplitude of such currents is reduced (compared to the CAR amplitude in the interval $[ -\epsilon_1, \epsilon_1 ]$) 
because both DAR and EC require one electron transfer through a ``small'' resonance of $\rho_{1}$ and/or $\rho_{2}$.   

\begin{figure}[ht]
	\centering
		\includegraphics[width=10cm]{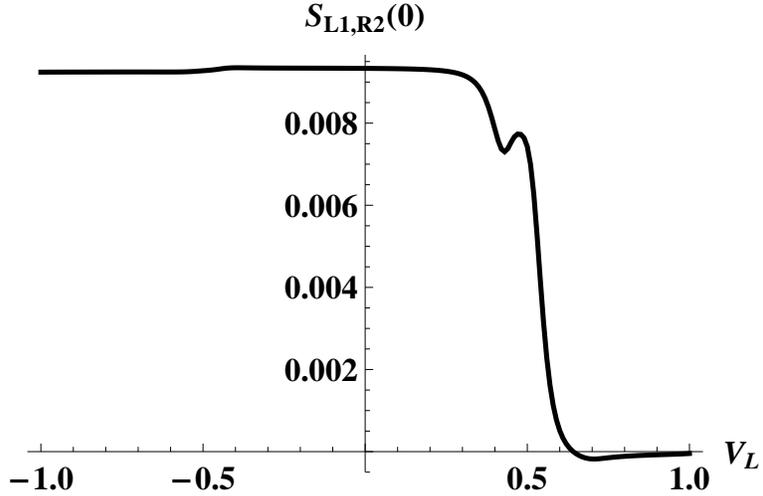}
	\caption{Current-current crossed correlation $S_{L1,R2}(0)$ (between $I_{L1}$ and $I_{R2}$) at zero frequency (arbitrary units) as a function of the voltage of the left lead $V_L$ for the same setup as in Fig. \ref{fig:Courant2}.}
	\label{fig:Bruit2}
\end{figure}

In order to confirm these observations, we plot in Fig. \ref{fig:Bruit2} the current-current crossed correlation at zero frequency between $I_{L1}$ and $I_{R2}$. The crossed correlation is positive until $\epsilon_1$ where the signal drops to zero and becomes eventually negative. Strictly speaking there is a small structure at $-\epsilon_1$ due to the (small) resonance of the density of states. 
At $\epsilon_1$ the cross correlation has a small secondary peak which originates from the (large) double peak resonance
in the dot density of states. For voltages larger than $\epsilon_1$, the negative signal has a much weaker 
amplitude than the positive one (below $\epsilon_1$) because the associated currents are smaller.  

We now give a physical interpretation of these results. When the voltage is smaller than $\epsilon_1$, the dominant process 
is CAR and  the crossed correlations are positive because the two electrons of the same Cooper pair are split and 
end up in opposite leads. When the voltage is larger than $\epsilon_1$ the dominant process is EC and the crossed correlations are negative. An electron injected from the left lead, which tunnels trough the superconductor then enters the right lead.
Note that the generated currents by DAR do not contribute to the crossed correlation because the two Cooper pairs which are injected in each side of the superconductor are independent. 

\subsubsection{Symmetric case} 

\begin{figure}[ht]
	\centering
		\includegraphics[width=8cm]{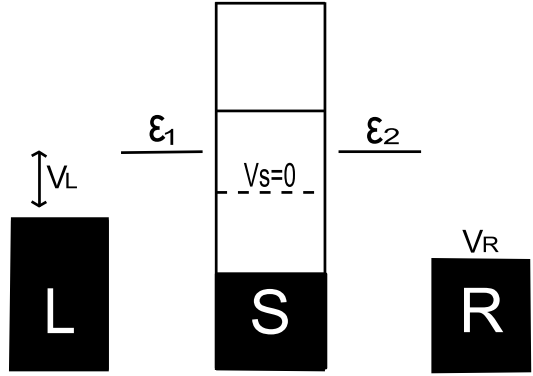}
	\caption{Symmetric case: energy level of the dots are the same. The voltage of the right lead $V_R$ is fixed and we vary the voltage of the left one $V_L$.}
	\label{fig:Sym}
\end{figure}

We now focus on the second geometry where the two dots levels have the same position (see Fig. \ref{fig:Sym}). Again we fix the voltage of the right lead $V_R$ and we vary the voltage of the left one $V_L$.

\begin{figure}[ht]
	\centering
		\includegraphics[width=10cm]{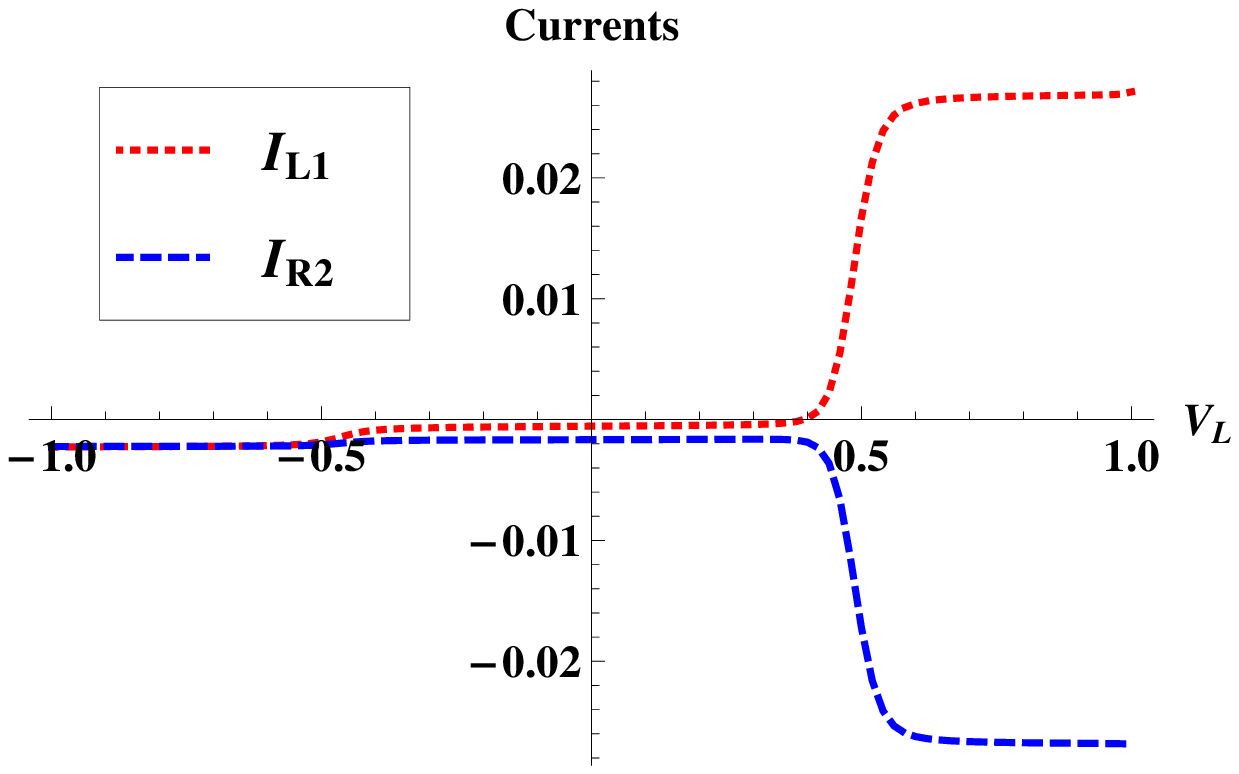}
	\caption{(Color online) Currents (arbitrary units) as a function of the voltage of the left normal lead $V_L$ for $\epsilon_1=0.5$, $\epsilon_2=0.5$, $V_R=-0.7$, $\beta=100$, $t_{L1}=t_{R2}=t_{S1}=t_{S2}=0.2$ and $t_{L2}=t_{R1}=0$.}
	\label{fig:Courant3}
\end{figure}

\begin{figure}[ht]
	\centering
		\includegraphics[width=10cm]{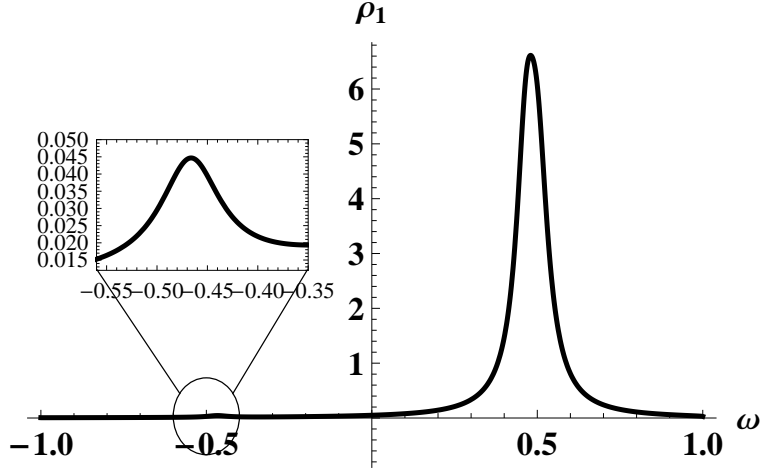}
	\caption{Density of states of the dot 1 (arbitrary units) for the same setup as in Fig. \ref{fig:Courant3}.}
	\label{fig:Density0sym}
\end{figure}

In Fig. \ref{fig:Courant3} we plot the currents in the two normal leads. We see that contrary to the previous case, the two currents 
have a reduced amplitude below $V_L=\epsilon_1$ and this amplitude increases above this voltage. 
To be more precise, the two currents are negative and identical below $V_L=-\epsilon_1$.
In the interval $[ -\epsilon_1, \epsilon_1 ]$ the currents remain negative but they differ slightly.
Above $V_L=\epsilon_1$, these currents bear opposite sign and their amplitude has increased.

When the voltage of the left lead is smaller than $-\epsilon_1$, the dominant processes are CAR and DAR. These two processes
distribute electrons in the two leads in an equivalent manner, hence they are equal. The smallness of their amplitude can be 
explained as above from density of states considerations (see Fig. \ref{fig:Density0sym}). Note that in this geometry $\rho_{1}=\rho_{2}$: both contain a large peak
at $\epsilon_1$ and a much smaller one at $-\epsilon_1$. For both CAR and DAR processes, an electron transfer through the small resonance is required, which explains their amplitude. 
In the interval $[ -\epsilon_1, \epsilon_1 ]$, DAR processes from the left lead are suppressed altogether, but they still operate
with the right lead. The CAR process also contributes (the two currents still bear the same sign), and it involves electron transfer
through both the ``large'' resonance of dot 1 and ``small'' resonance of dot 2 (hence the reduced amplitude). 
Note that in addition there arises a small contribution from EC processes, but this contribution is less important 
because the electron has to be transfered through both ``small'' resonances of the dots.

When the voltage of the left lead is larger than $\epsilon_1$, the dominant process is EC because electrons of the left lead can be injected in the  first dot (through its ``large'' resonance) they tunnel trough the superconductor and then they end up
in the right lead via the second dot (through its ``large'' resonance also). In this configuration the DAR is always present but its 
amplitude is reduced, once again because of density of states arguments (passage through a ``small'' resonance). 
These two processes give currents with opposite sign because either electrons from the left lead go into the right lead (EC processes), or 
two electrons from the left lead form a Cooper pair in the superconductor (DAR from left lead), or Cooper pairs are injected from the superconductor into the right lead (DAR into right lead). CAR is killed because injection of electrons in the left lead is prohibited since the voltage is larger than the dot 1 energy. The amplitude of the generated currents in this configuration is important because this corresponds to the optimal configuration for EC. Electron transfer then exploits the maximum of the density of states for both dots.

\begin{figure}[ht]
	\centering
		\includegraphics[width=10cm]{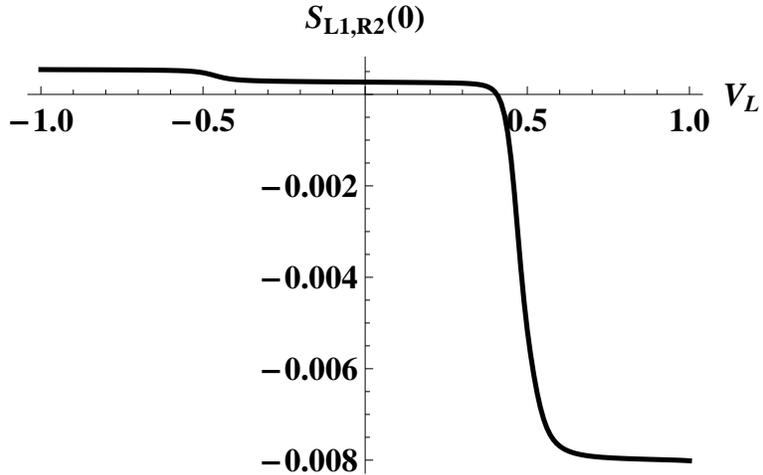}
	\caption{Current-current crossed correlation $S_{L1,R2}(0)$ (between $I_{L1}$ and $I_{R2}$) at zero frequency (arbitrary units) as a function of the voltage of the left lead $V_L$ for the same setup as in Fig. \ref{fig:Courant3}.}
	\label{fig:Bruit3}
\end{figure}

In Fig. \ref{fig:Bruit3} we plot the current-current crossed correlation at zero frequency between $I_{L1}$ and $I_{R2}$. The correlations are positive below $\epsilon_1$ then become negative, with a large amplitude. Below $\epsilon_1$ the amplitude of the signal is reduced and it has a structure at $V_L=-\epsilon_1$. 

When the voltage is smaller than $\epsilon_1$, the dominant process is CAR thus the crossed correlations are positive and the explanation is the same as in the anti-symmetric case. The reduced amplitude of the crossed correlation is explained by the fact that one electron has
to be transfered through a ``small'' resonance. The feature at $V_L= -\epsilon_1$ corresponds to the onset for EC process through the
small resonances of the two dots. Above $V_L= \epsilon_1$ electrons can pass through both large resonances of the dot which explains the large (negative) amplitude of the signal. The generated currents by DAR don't contribute to the crossed correlation because the two Cooper pairs which are injected in each side of the superconductor are independent.

To summarize we can facilitate the EC regime if we have the same energy level position for the two dots (Symmetric case) and the CAR regime is facilitated if we have opposite energy level for the two dots (Anti-symmetric case). In the following section we are going to be interested in the effect of direct tunneling between the two dots, which is relevant in recent experiments.\cite{herrmann,hofstetter}

\subsection{Transport with tunneling between the dots}

In this section we allow tunneling between the two dots ($t_d\neq0$) and we study the effect of this manipulation on the currents and the crossed correlation at zero frequency.

\subsubsection{Anti-symmetric case}

\begin{figure}[ht]
	\centering
		\includegraphics[width=10cm]{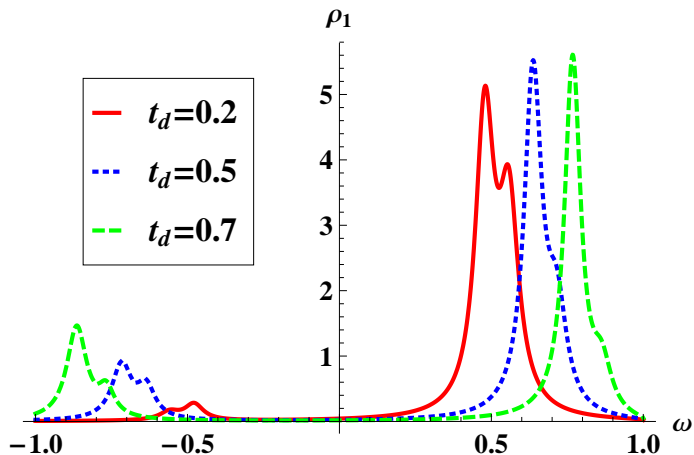}
	\caption{(color online) Density of states of the dot 1 (arbitrary units) for various values of the tunneling between the two dots and for $\epsilon_1=0.5$, $\epsilon_2=-0.5$, $t_{L1}=t_{R2}=t_{S1}=t_{S2}=0.2$ and $t_{L2}=t_{R1}=0$.}
	\label{fig:DensityAnt}
\end{figure}

The main effect associated with tunneling between the dots is that it modifies the density of states of each dot. In Fig. \ref{fig:DensityAnt} we see that for the anti-symmetric case, the weight of the small resonance (at $-\epsilon_1$ for dot 1, and at $\epsilon_1$ for dot 2) is increased by such tunneling. The large resonance (at $\epsilon_1$ for dot 1, and at $-\epsilon_1$ for dot 2) now acquires an asymmetric double 
peak structure, for positive energies, which disappears for sufficiently strong tunneling between the dots.

\begin{figure}[ht]
	\centering
		\includegraphics[width=10cm]{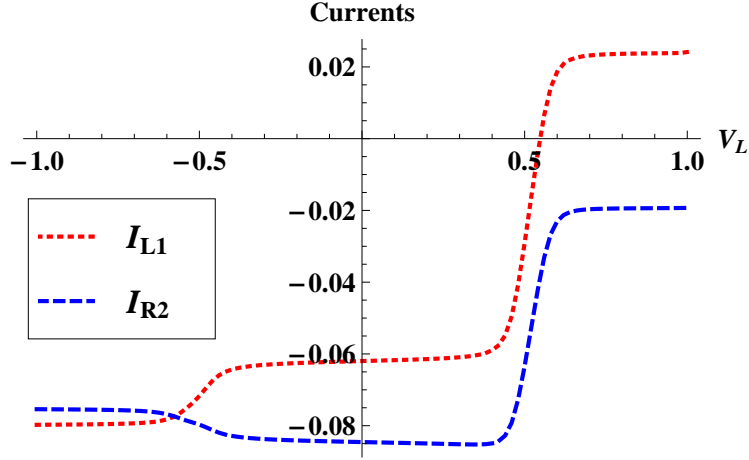}
	\caption{(color online) Currents (arbitrary units) as a function of the voltage of the left normal lead $V_L$ for $\epsilon_1=0.5$, $\epsilon_2=-0.5$, $V_R=-0.7$, $\beta=100$, $t_{L1}=t_{R2}=t_{S1}=t_{S2}=0.2$, $t_{L2}=t_{R1}=0$ and the tunneling between the dots $t_d=0.2$.}
	\label{fig:Antiope}
\end{figure}

The two currents are plotted in Fig. \ref{fig:Antiope}. For $V_L<-\epsilon_1$, both currents are negative with a large amplitude associated to CAR, as in section \ref{anti_symmetric_no_tunneling}. Contrary to section \ref{without_tunneling}, they have a different amplitude
($I_{R2}>I_{L1}$). There are two dominant processes into play. On the one hand, the presence of tunneling between the dots allows to transfer electrons from right to left lead without passing through the superconductor. On the other hand, as the (small) resonance of dot 1 has been increased, there is also now the possibility for EC from dot 2 to dot 1 (this effect was not noticeable in the absence 
of tunneling between the dots). In addition, there is also the possibility for DAR process which injects electrons in both leads.
 
Increasing $V_L$ beyond $-\epsilon_1$, CAR is still dominant, but the currents cross ($I_{L1}>I_{R2}$ as in section \ref{anti_symmetric_no_tunneling}) because both direct tunneling and EC from dot 1 to dot 2 come into play. The difference $I_{L1}-I_{R2}$ is larger than in the absence of tunneling between dots because of the increased weight of the small resonance in the density of states. 
In this situation, there is also a contribution for DAR processes in the right lead. 

Finally, for $V_L>\epsilon_1$ CAR is suppressed. In this regime, the main processes are direct tunneling 
and EC from dot 1 to dot 2, as well as DAR from both the left lead to the superconductor, and the superconductor to the right lead. The difference $I_{L1}-I_{R2}$ is further increased by 
both direct tunneling between dots and the increase of EC associated with density of states effects.  

\begin{figure}[ht]
	\centering
		\includegraphics[width=12cm]{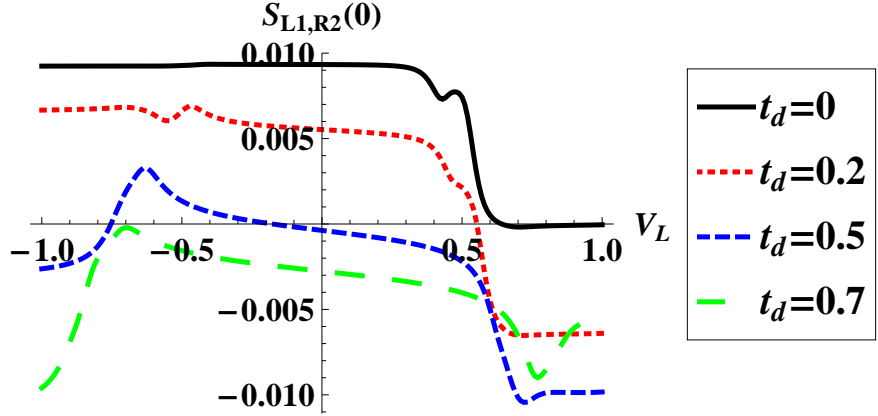}
	\caption{(color online) Current-current crossed correlation $S_{L1,R2}(0)$ (between $I_{L1}$ and $I_{R2}$) at zero frequency (arbitrary units) as a function of the voltage of the left lead $V_L$ for various values of the tunneling between the two dots and for $\epsilon_1=0.5$, $\epsilon_2=-0.5$, $V_R=-0.7$, $\beta=100$, $t_{L1}=t_{R2}=t_{S1}=t_{S2}=0.2$ and $t_{L2}=t_{R1}=0$.}
	\label{fig:NAntop}
\end{figure}

In Fig. \ref{fig:NAntop} we plot the zero frequency current-current crossed correlation as a function of the voltage of the left normal lead for various values of the tunneling between the two dots. For sufficiently weak tunneling parameters, the general tendency is to favor positive correlations for $V_L<\epsilon_1$ and negative cross correlations for $V_L>\epsilon_1$, signaling a transition from CAR processes to EC processes. When the tunneling between dots is increased, the crossed correlations are shifted towards negative values, which results in 
a reduced positive signal in the CAR regime and an increased (negative) signal for the direct tunneling and EC regime. Beyond $t_d=0.7$, the positive 
crossed correlations signal disappear completely. 

We further discuss the structure associated with the dot resonances at $\pm \epsilon_1$. Increasing the direct tunneling between the dot
affects their density of states (see Fig. \ref{fig:DensityAnt}). As mentionned above, the ``large'' resonance at $\pm \epsilon_1$ (for dot 1/dot 2) has a double peak structure which disappears when $t_d$ is increased, while the ``small'' resonance acquires a double peak structure when increasing $t_d$. 
This constitutes the justification for the side peak in the current-current correlation near $V_L=\epsilon_1$ to be smoothed out
in the presence of strong tunneling. At the same time, for intermediate tunneling $t_d=0.2$, the new double peak structure 
in the density of states of dot 1 at $V_L=-\epsilon_1$ creates a structure in the (positive) current crossed correlation. Further increasing $t_d$, 
a peak in the crossed correlation around $V_L=-\epsilon_1$ is generated: this peak is shifted towards  $V_L<-\epsilon_1$ in accordance
with the shift in energy which is observed in the density of states of dot 1. This peak is the last stronghold for the observation 
of positive crossed correlations: for voltages below (above) it, direct tunneling between the dots transfers electrons from the right to the left lead
(from the left to the right lead). We therefore interpret the presence of this peak as the point where the main process in competition with CAR, which is the direct tunneling, changes sign.  

\subsubsection{Symmetric case}

\begin{figure}[ht]
	\centering
		\includegraphics[width=10cm]{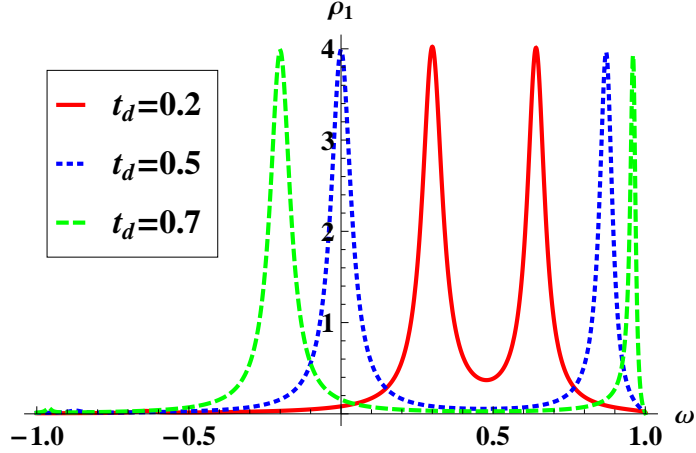}
	\caption{(color online) Density of states of the dot 1 (arbitrary units) for various values of the tunneling between the two dots and for $\epsilon_1=0.5$, $\epsilon_2=0.5$, $t_{L1}=t_{R2}=t_{S1}=t_{S2}=0.2$ and $t_{L2}=t_{R1}=0$.}
	\label{fig:Densitysym}
\end{figure}

The density of states (which is the same for dot 1 and dot 2) is drastically different in the case where the two dots have the same level position when the tunneling between dots is switched on (see Fig. \ref{fig:Densitysym}). For moderate tunneling between the dots, it contains a (large) double peak at positive energies which is centered near $\epsilon_1$, and a much smaller peak (not shown) close to $-\epsilon_1$. Further increasing the tunneling, the double peak becomes well separated, with one side peak shifted towards negative energies, while the other side peak approaches the superconducting gap. 

\begin{figure}[ht]
	\centering
		\includegraphics[width=10cm]{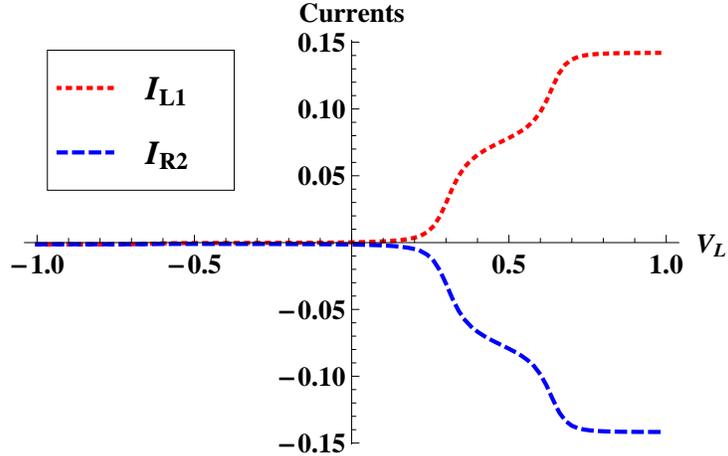}
	\caption{(color online) Currents (arbitrary units) as a function of the voltage of the left normal lead $V_L$ for $\epsilon_1=0.5$, $\epsilon_2=0.5$, $V_R=-0.7$, $\beta=100$, $t_{L1}=t_{R2}=t_{S1}=t_{S2}=0.2$, $t_{L2}=t_{R1}=0$ and the tunneling between the dots $t_d=0.2$.}
	\label{fig:Symope}
\end{figure}

The two currents are plotted in Fig. \ref{fig:Symope}.
For $t_d=0.2$ the currents $I_{L1}$ and $I_{R2}$ remain negative for $V_L<0$, which is symptomatic of CAR and DAR processes, 
but their amplitude is drastically reduced because of the lack of weight in the density of states for negative
energies. For $V_L>0$, the two currents acquire an opposite sign and we observe two steps which correspond 
to the double peak structure in the density of states. This corresponds to the situation where both 
direct tunneling and EC processes contribute in the same direction. Further increasing the coupling between 
the dots enhances these features.  

\begin{figure}[ht]
	\centering
		\includegraphics[width=12cm]{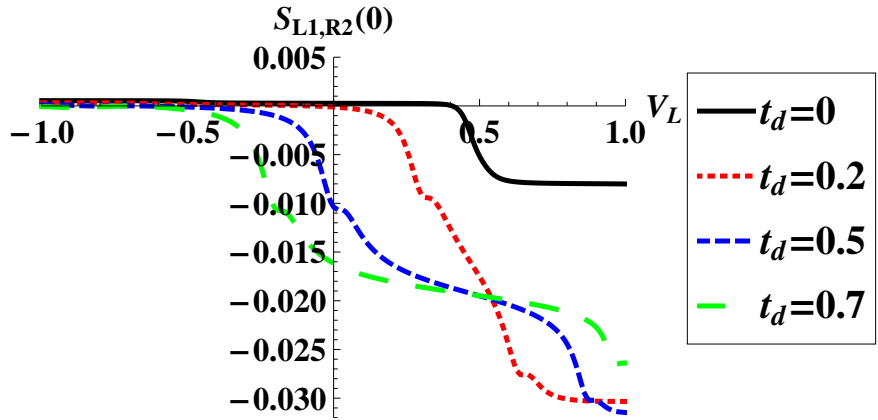}
	\caption{(color online) Current-current crossed correlation $S_{L1,R2}(0)$ (between $I_{L1}$ and $I_{R2}$) at zero frequency (arbitrary units) as a function of the voltage of the left lead $V_L$ for various values of the tunneling between the two dots and for $\epsilon_1=0.5$, $\epsilon_2=0.5$, $V_R=-0.7$, $\beta=100$, $t_{L1}=t_{R2}=t_{S1}=t_{S2}=0.2$ and $t_{L2}=t_{R1}=0$.}
	\label{fig:NSymop}
\end{figure}

The crossed correlations are plotted in Fig. \ref{fig:NSymop} for several values of the tunneling
amplitude between the dots. The sign is positive but vanishingly small for $V_L<0$, which is again the consequence
of the drastically reduced density of states for negative energies. The cross correlations decrease monotonously
when $V_L$ is increased, but unlike the case of zero coupling between the dots the crossover from the CAR 
dominated regime to the EC dominated regime now occurs close to $V_L=0$ at $t_d=0.2$. The noise cross correlation acquires an
appreciable amplitude when $V_L$ is increased beyond the characteristic energies associated with the 
double peak structure in the density of states (typically $V_L= 0.3$ and $V_L=0.7$ for $t_d=0.2$).
At the location of these peaks one observes some features in the crossed correlation signal. For larger tunneling 
between the dots this monotonous decrease of the cross correlation, and its (larger) amplitude is even 
more pronounced. It can even occur for $V_L<0$ because the density of states of both dots acquires a peak
at negative energies, thus favoring EC process as well as direct tunneling.    

\section{Conclusion}
\label{sec:conclusion}

To summarize, we have developped a framework for the description of the transport properties of
a device consisting of a superconducting finger connected to two normal metal leads via two
quantum dots adjacent to the superconductor. Two recent experiments \cite{hofstetter,herrmann} 
measured the branching currents in such a device, and showed that non local effects were at play, 
which is consistent with Cooper pair splitting. Yet it is now well accepted that the evidence of Cooper 
pair splitting in such a device would be more robust if current-current crossed correlations were measured, which
is the main justification for the present work. 

For pedagogical reasons, we started our analysis with
the description of a device where the superconductor is connected to a single dot 
(which is in turn connected to normal metal leads), in order to 
derive the self-energy which arises from the coupling of the dot to all leads.
Next, we focused on the experimental geometry of Refs. \onlinecite{hofstetter} and \onlinecite{herrmann} with
two dots, allowing for direct tunneling between the two. The current and noise crossed correlation were computed using 
the Keldysh formalism. The branching currents were expressed in terms of the single particle Green's
function of the dot, which is dressed by the coupling to the leads, in the same spirit as the 
Fisher-Lee formula\cite{fisher_lee}. Similarly, the noise (and in particular the noise crossed correlations)
was expressed in terms of a two particle (dressed) Green's function for the dot variables. This 
second result corresponds to an extension of the Fisher-Lee/Landauer-B\"uttiker formula for the noise. 
Provided a given choice of interactions within the dot (Coulomb, electron-phonon,...)
these formulae for the currents and current-current crossed correlation are exact if the single and two particle Green's functions can
be computed. For our purposes, we chose to illustrate the operation of this device 
by ignoring such interactions. In this case the two particle Green's function can be decoupled as 
a sum of products of single particle Green's functions. 

The current and the noise crossed
correlation were obtained numerically by solving the Dyson's equation. Given the complexity of the 
device, the number of parameters (voltages, energy levels, resonance widths) constrained us 
to focus on few specific cases. We focused on transport for voltages and energy levels all contained within 
the superconducting gap, because this corresponds to the regime where Cooper pair splitting is 
understood to occur. We looked at two configurations for the dot energy levels: the antisymmetric
configuration (the two levels have opposite energies with respect to the superconducting chemical potential)
and the symmetric configuration (same energies for the two levels). 
A crucial concept for the understanding of the generated data for current and noise is the density 
of states of the dots: due to the proximity effect induced by the presence of the superconductor, 
the density of states of a given dot acquires some weight (``small'' resonance) at the energy which is opposite
to the resonant energy of the bare dot (``large'' resonance). Electron transport is therefore favored if the 
electrons can choose to be transfered via either a ``large'' or a ``small'' resonance of these 
dots. We fixed the voltage of a given lead within the gap, below all resonances, and we varied 
the voltage of the other normal metal lead within the whole gap. 
We started the discussion assuming no direct coupling between the two dots. On one hand we observed that 
CAR process is optimized in the antisymmetric case, when both normal lead voltages are below the 
dot density of states resonances. On the other hand we observed that EC process 
is the most important one in the symmetric configuration, when the voltage which is varied 
is set above all resonances. A systematic study of the current/voltage characteristic
and of the noise/voltage characteristic allowed to identify precisely which processes (CAR, DAR, EC) are into play 
when the voltage is increased. We argued that both the monitoring of the branching currents as well
as the noise crossed correlations are necessary to identify these processes: for instance, 
DAR processes contribute to the current, but not to the noise correlation signal. 
Finally, we switched on the direct coupling between the dots and we repeated the analysis. 
The density of states of the dots undergoes strong modifications in the presence of such a 
coupling. This coupling has a tendency to spoil the positive correlations: for sufficiently strong 
coupling, positive correlations disappears altogether. We therefore provided a rather complete picture 
of the operation of this Cooper pair beam splitter which could be useful in future noise crossed correlation
experiments. 

Several extensions of this work can be envisioned. First, we clearly neglected the separation between 
the two injection locations of the superconductor/dot interfaces. In all theoretical and experimental
investigations, this separation $r$ has to be smaller than the superconducting coherence length (the ``size''
of the Cooper pair in the superconductor), because this leads to an exponential decay of the CAR (and EC)
process. In addition to this reduction, a power law decay with $k_Fr$ (which depends on the dimensionality) 
is typically found theoretically,\cite{sukhorukov_recher_loss,falci_feinberg_hekking,melin_feinberg_europhys,melin_feinberg}  and it can lead to a strong suppression of CAR and EC processes. Ref. \onlinecite{hofstetter} did not find conclusive evidence of this power law suppression.
It argued that the segment of the nanowire which is buried below the superconductor, which acquires
a minigap, plays the dominant role for Cooper pair splitting and should be immune to the power law decay
because electrons are directly injected to the dots on both sides. 
Nevertheless, further investigations of these power law suppression with our setup could provide additional insight.    
 
Second, we have neglected the Coulomb interactions within the dot, which is justified if the 
resonant linewidth is sufficiently large (but not larger than the superconducting gap) so 
as to constitute open quantum dots. The inclusion of interactions within the dots 
constitutes a definite challenge, and such interactions can in practice only be included
via approximate treatments such as mean field theory, perturbative diagrammatic resummation, 
or Kondo phenomenology. In practice, electron interactions allow to transfer electrons one by one
through each dot. In our device, we can only argue that successive Cooper pairs 
emitted by the superconductor do not overlap significantly if the boundary between the 
superconductor and the dots is sufficiently opaque. 
In the two experiments which are relevant to this work, it was possible 
to characterize the dots by switching off the superconductivity by applying a magnetic field. 
Such analysis clearly showed the presence of Coulomb diamonds, and the boundaries between them 
allowed to identify parameters for resonant electron transfer. 
Here we can only argue that 
regardless of the physical origin of such resonances, they will have a characteristic location and width
associated with interactions, and they can be effectively  
tuned by adjusting the gate voltages on the two dots. The scenario for adjusting 
them antisymmetrically or symmetrically to favor CAR/EC processes should then be robust.
But given the fact that our intermediate results for the current and noise can be 
expressed in terms of exact one and two particle Green's functions, an extension of the present work
including interactions using an approximate scheme is foreseeable in the future.     

Finally, we have focused solely on the zero frequency noise crossed correlation signal. Noise crossed 
correlations at finite frequencies contain additional information, such as the relevant time scale
for Cooper pair splitting in this particular device. Several entanglement scenarios have 
addressed the importance of short times/high frequencies, for instance when performing
an entanglement diagnosis.\cite{lebedev,bayandin_energy}
Further investigations along these lines would prove 
useful. 

\textit{Note added in proof}. Upon completion of the manuscript, we failed to include some relevant references on master equations type approaches which are applied to study transport in the entangler of Ref. \onlinecite{sukhorukov_recher_loss}. Refs. \onlinecite{sauret_feinberg_martin} and \onlinecite{governale_konig} derived the quantum master equation for the entangler and computed the branching currents in this device. Ref. \onlinecite{governale_konig} focused on the experimentally measurable quantities in the light of the recent experiments Refs. \onlinecite{hofstetter} and \onlinecite{herrmann}. Ref. \onlinecite{sauret_feinberg_martin_2} computed the zero frequency noise crossed correlator in order to achieve a Bell inequality test within this approach.

\acknowledgments

We acknowledge fruitful discussions with  T. Kontos and E. Paladino.

\end{document}